\documentclass[12pt,preprint]{aastex}
\begin{document}
\title{A re-visit of the phase-resolved X-ray and $\gamma$-ray spectra of the Crab pulsar}

\author{Anisia P.S. Tang}
\affil{Department of Physics, The University of Hong Kong, Pokfulam Road, Hong Kong, P.R. China}
\author{J. Takata}
\affil{ASIAA/National Tsing Hua University, TIARA, Hsin-Chu, Taiwan}
\author{J.J. Jia}
\affil{Department of Physics and Astronomy, Johns Hopkins University, Baltimore, MD21218, USA}
\and
\author{K.S. Cheng}
\affil{Department of Physics, The University of Hong Kong, Pokfulam Road, Hong Kong, P.R. 
China}


\begin{abstract}
We use a modified outer gap model to study the multi-frequency phase-resolved spectra of 
the Crab pulsar. The emissions from both poles contribute to the light curve and the phase-resolved spectra. Using the synchrotron self-Compton mechanism and by considering the incomplete conversion of curvature photons into secondary pairs, the observed phase-averaged spectrum from $100$ eV - $10$ GeV can be explained very well. The predicted phase-resolved spectra can match the observed data reasonably well, too. We find that the emission from the north pole mainly contributes to Leading Wing 1. The emissions in the remaining phases are mainly dominated by the south pole. The widening of the azimuthal extension of the outer gap explains Trailing Wing 2. The complicated phase-resolved spectra for the phases between the two peaks, namely Trailing Wing 1, Bridge and Leading Wing 2, strongly suggest that there are at least two well-separated emission regions with multiple emission mechanisms, i.e.~synchrotron radiation, inverse Compton scattering and curvature radiation. Our best fit results indicate that there may exist some asymmetry between the south and the north poles. Our model predictions can be examined by GLAST.
\end{abstract}




\section{Introduction}

It is generally believed that the phase-resolved spectra can provide the most detailed information about the structure of the pulsar magnetosphere, acceleration mechanism, and, pair creation and radiation processes in the outer magnetosphere. \citet{a31} and \citet{DH96} firstly calculated the phase-resolved spectra of the Vela pulsar. Then, \citet{a1} and \citet{ZC02} calculated the phase-resolved spectra of the Crab pulsar in the $\gamma$-ray and the X-ray regimes separately. It will be interesting to find a general scenario that can produce both the $\gamma$-ray and the X-ray regimes in the phase-resolved spectra of the Crab pulsar. Our preliminary results have been reported in the 2006 COSPAR conference \citep{COSPAR07}. We have continued to study the phase-resolved spectra of the Crab pulsar in more details. Here, we present our recent results with a different parametric fitting \citep{myphdthesis}.

According to \citeauthor{CHRa} (\citeyear{CHRa,CHRb}; hereafter CHR), the outer gap starts at the null charge surface, ends at the light cylinder, is bounded below by the last closed field line and is bounded above by a layer of electric current which replenishes charges to the open field lines outside the gap to maintain a steady charge density, the Goldreich-Julian charge density \citep{a27},
\begin{equation}
\rho _{\rm GJ}\sim-\frac{\mbox{\boldmath$B$}\cdot\mbox{\boldmath{$\Omega$}}}{2\pi c}.
\end{equation}
The charge depletion within the outer gap due to global flows of charged particles causes a large electric field along the magnetic field lines so that $\mbox{\boldmath$E$}\cdot\mbox{\boldmath$B$}\neq 0$ inside the gap. This gap thus acts as an accelerator to boost the charged particles to relativistic speeds. Through a cascade process, high energy $\gamma$-ray photons and e$^\pm$ pairs are produced. Recently, the classical outer gap model is being challenged by \citet{HHS03}. By solving the set of Maxwell and Boltzmann equations, they find that a current at nearly Goldreich-Julian rate can shift the position of the inner boundary of the outer gap. Therefore, we adopt a modified version of the CHR outer gap model such that the inner boundary of the outer gap is shifted inwards.

The photon emission mechanism starts with curvature radiation of the accelerated charged particles in the gap. The emission direction is tangent to the local magnetic field lines. As a photon escapes, it may encounter a low energy photon. The low energy photon may be a thermal photon from the stellar surface or a magnetospheric soft synchrotron photon emitted by the secondary e$^\pm$ pairs which are created by the curvature photons from the inner field lines. The primary curvature photons will then be converted into the secondary e$^\pm$ pairs via photon-photon pair production, i.e.~$\gamma+\gamma\rightarrow{\rm e}^++{\rm e}^-$ or $\gamma+X\rightarrow{\rm e}^++{\rm e}^-$ inside and outside the gap. As pointed out in \citet{a1}, although pair production inside an outer gap is limited to a small region, pair production outside the outer gap can cover a much wider area because the synchrotron photons produced by the secondary pairs are more abundant than the thermal photons from the stellar surface. Although the secondary synchrotron photons cannot get into the outer gap due to the field line curvature, they can convert most of the primary curvature photons from the outer gap into secondary pairs. Since the synchrotron radiation is beamed to a small angle by the relativistic beaming effect \citep{radiative}, synchrotron photons will also be seen as more or less tangent to the field lines if observed.

Besides curvature radiation and synchrotron radiation, inverse Compton scattering is another important radiation mechanism in the neutron star magnetosphere. It occurs when a fast moving electron or positron collides with a photon and net energy is transferred from the particle to the photon. In the far region of the magnetosphere, the relativistic particles collide with the soft synchrotron photons through the inverse Compton scattering process.

\citet{CR92,CR94} and \citet{a29} calculated the light curves by considering a single outer gap with the photon emission to be beamed to the outside alone in order to avoid the multiple-peak feature that does not occur in true observation data of the Crab pulsar. \citet{a1} pointed out the lack of a reason to explain the ignorance of the incoming photons (the photons that are beamed towards the star). However, in Section~\ref{emigeo}, we estimated that inward emission is much fainter than the outward emission (see also \citeauthor{a1} \citeyear{a1}). Furthermore, since the previous models by \citet{a29} and \citet{a1} considered emission beyond the null charge surface only, they suggest that the observer can measure the photons from one gap only, that is, a single-pole model is considered in the canonical model. However, if the gap extends below null charge surface, the photons originating from another gap become measurable by the observer as well. In this paper, therefore, although we follow \citet{CR92,CR94} and \citet{a29} to calculate the light curve, we consider both gaps, that is, a two-pole model is examined.

In this paper, we calculate the pulse profile and the phase-resolved spectra for the Crab pulsar with an outer gap accelerator model. Because the Crab pulsar is one of the brightest $\gamma$-ray sources in the sky, the detailed observation for the pulse profile and the phase-resolved spectra have been obtained. These provide useful information to the study of the non-thermal processes in the pulsar magnetosphere. Since the Crab pulsar is still young and is believed to have a thin gap, calculating the contribution from one layer suffices. By fixing an inclination angle and an observer angle, the photon emission locations that produce the light curve are used to calculate the phase-resolved photon spectra which are compared with the observation data. For the Crab pulsar, we adopt $R=10$~km for the stellar radius and $B_p=3.8 \times 10^{12}$~Gauss for the stellar magnetic field strength.

Another major feature in this paper is the relaxing of the assumption that all curvature radiation has been converted into e$^\pm$ pairs as in the previous studies. In fact, by letting trace amount of the curvature radiation, which is emitted far away from the star, to survive and escape, we can obtain better fitting curves for the phase-resolved spectra of the Crab pulsar, especially in the phases of Trailing Wing 1, Bridge and Leading Wing 2.

In Section~\ref{section:lightcurve}, we review briefly the emission geometry in the magnetosphere. Then we discuss the location of the inner boundary of the outer gap and produce the light curve for the Crab pulsar. In Section~\ref{section:spectra}, we discuss the electric field component along the magnetic field lines and the three major emission mechanisms, namely synchrotron radiation, inverse Compton scattering and curvature radiation. We argue that although most of the primary curvature photons have been converted to secondary pairs to produce synchrotron photons, some curvature photons which are emitted far away from the star can survive the cascade process and escape. They contribute to the peak in the high energy regime at several GeV of the phases Trailing Wing 1, Bridge and Leading Wing 2. In the final section, we conclude our results and discuss briefly about the justification of our assumptions.

\section{Theoretical light curves of neutron stars with Crab parameters\label{section:lightcurve}}

\subsection{Emission geometry}
\label{emigeo}
To calculate the light curves and the spectra, we adopt the rotating dipole field in the magnetosphere. For a rotating dipole, the local magnetic field $\mbox{\boldmath$B$}(r)$ is given by \citep{a1}
\begin{equation}
\mbox{\boldmath$B$}=\hat{r}\left[\hat{r}\cdot\left(\frac{3\mbox{\boldmath$\mu$}
}{r^3}
+\frac{3\dot{\mbox{\boldmath$\mu$}}}{cr^2}+\frac{\ddot{\mbox{\boldmath$\mu$}}}{c^2r}
\right)\right]-\left(\frac{\mbox{\boldmath$\mu$}}{r^3}+\frac{\dot{\mbox
{\boldmath$\mu$}}}{cr^2}+\frac{\ddot{\mbox{\boldmath$\mu$}}}{c^2r}\right),\nonumber
\label{eq:Br}
\end{equation}
where $\mbox{\boldmath$\mu$}=\mu\left(\sin\alpha\cos\Omega t\hat{x}+\sin\alpha\sin\Omega t\hat{y}+\cos\alpha\hat{z}\right)$ is the magnetic moment vector, $\hat{r}$ is the radial unit vector and $\alpha$ is the inclination angle.

\begin{figure}
\epsscale{1} \plotone{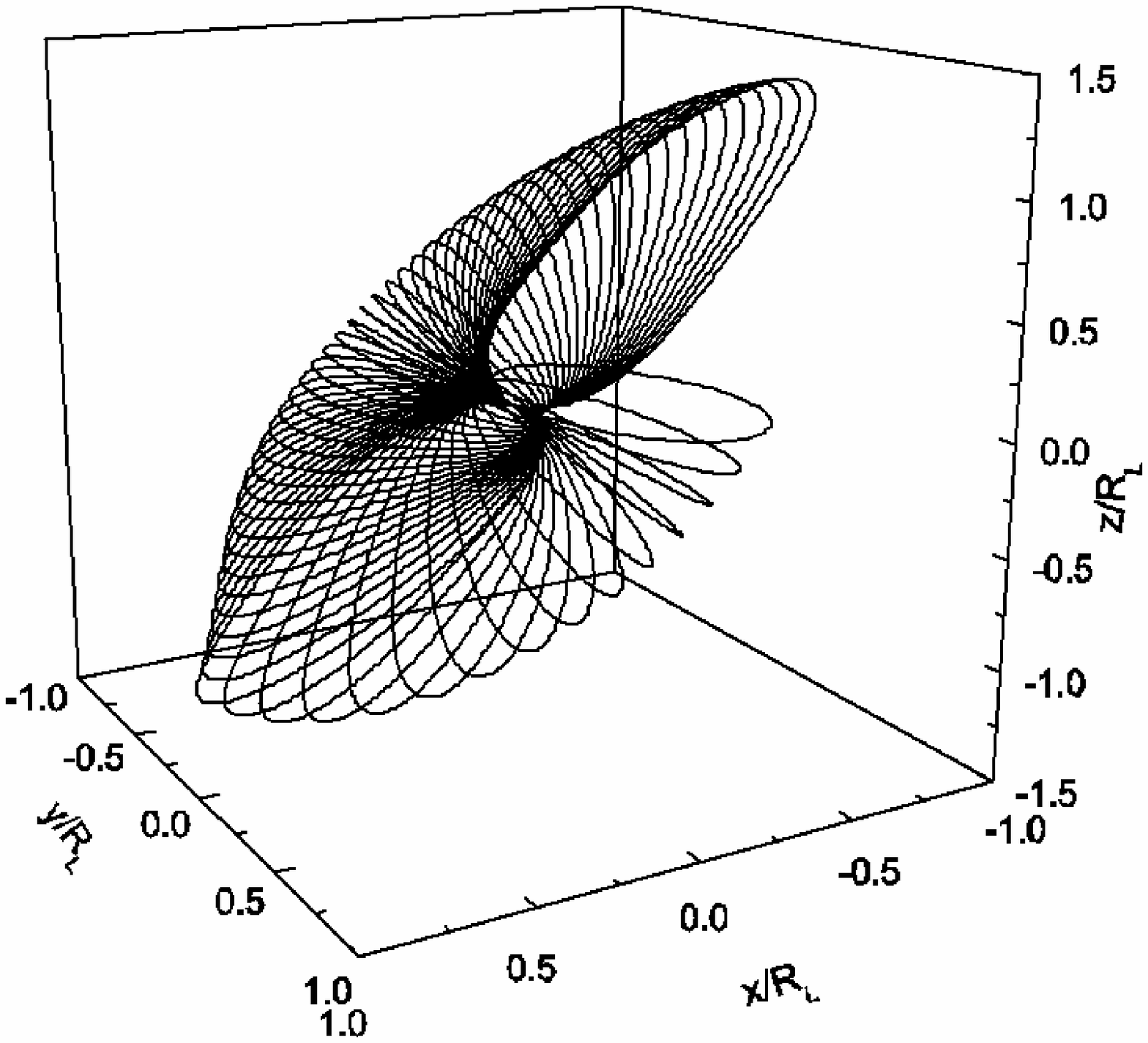}
 \caption{The 3-dimensional view of the
last closed field lines for $\alpha=50^\circ$ in the rotating frame.
The rotational axis is the $z$-axis. The
$\mbox{\boldmath$\Omega$}$-$\mbox{\boldmath$\mu$}$ plane is the
$x$-$z$ plane. Crab period is used: $P=0.33$ s, so $R_L\sim
159R\approx 1.59\times 10^8$ cm.\label{fig:3d}}
\end{figure}

We calculate the polar cap edge by using $(x_0,y_0,z_0)=(R_p\cos\phi _p,R_p\sin\phi _p,\sqrt{R^2-R_p^2})$ as the initial trial values for the computer program. $R_p=R\sqrt{R/R_L}$ is the polar cap radius of an aligned static dipole where $R$ is the stellar radius and $R_L=c/\Omega$ is the radius of the light cylinder. $\phi _p$ is the azimuthal angle about the magnetic axis. We call it the polar cap angle. Then we employ the Runge-Kutta method to trace out the field lines. In subsequent iterations, we try to find the scaling factors, $a_0$, such that $(x_0',y_0',z_0')=(a_0x_0,a_0y_0,\sqrt{R^2-a_0^2R_p^2})$ correspond to the footprints of the last closed field lines, i.e.~the boundary of the polar cap. In Figure~2 of \citet{a1}, they have shown that $a_0$ is $\phi _p$-dependent. The polar cap of a rotating dipole is not circular in shape especially for those having large inclination angles. A three-dimensional view of the last closed field lines is shown in Figure~\ref{fig:3d}. Next we define another scaling factor $a_1$ to denote the footprints of the open field lines as $(x,y,z)=(a_1x_0',a_1y_0',\sqrt{R^2-(x^2+y^2)})$. Here $a_1=0$ represents the magnetic pole and $a_1=1$ represents the last closed field lines.

Since the star is rotating, aberration occurs along the line of sight. With $\beta=|\mbox{\boldmath$r$}\times\mbox{\boldmath$\Omega$}|/c$, we have
\begin{eqnarray}
u_\phi'&=&\frac{(u_\phi+\beta c)}{(1+\beta u_\phi/c)}\nonumber\\
u_\theta'&=&\frac{u_\theta\sqrt{1-\beta^2}}{(1+\beta u_\phi c)}\nonumber\\
u_r'&=&\frac{u_r\sqrt{1-\beta^2}}{(1+\beta u_\phi c)}\label{eq:aberration}
\end{eqnarray}
where $u_i$ and $u_i'$ for $i=r$, $\theta$ and $\phi$ are the emission direction in the co-rotating and observer frames, respectively. Choosing the rotational axis as the $z$-axis, in the observer frame, the polar angle from the rotational axis is given by \citep{yadthesis}
\begin{equation}
\cos\zeta=\frac{u_z'}{u'}.\label{eq:viewangle}
\end{equation}
where $\zeta$ is the viewing angle with $\zeta=0^\circ$ when the star is viewed directly above its rotational axis and $\zeta=90^\circ$ when the star is viewed over the stellar equator.

Comparing to the photons emitted at the centre of the star, a photon emitted at any particular location $\mbox{\boldmath$r$}$ will take less time to travel to the light cylinder. The phase difference due to the travel time is given by $\Delta\Phi=-\mbox{\boldmath$r$}\cdot\hat{u}'/R_L$. Therefore, we have the phase angle $\Phi$ in the observer frame \citep{yadthesis} to be given by
\begin{equation}
\Phi=-\phi'-\frac{\mbox{\boldmath$r$}\cdot\hat{u}'}{R_L}\label{eq:phase}
\end{equation}
where $-\phi'=-\cos^{-1}(u_x'/u_{xy}')$ is the azimuthal angle in the observer frame. Choosing \mbox{\boldmath$\Omega$}-\mbox{\boldmath$\mu$} plane to be the $x$-$z$ plane, $u_{xy}'$ is the length of the projection of $\hat{u}'$ on the $x$-$y$ plane.

 We would like to remark that we assume the dipole field in the corotating frame. 
On the other hand, \citet{Ta06} assumed  the rotating dipole field  
in the observer frame. With the difference in the magnetic field configuration 
in the observer frame, the emission direction near the light cylinder 
is azimuthal direction for the present case and nearly radial for the case in 
\citet{Ta06}.  Essentially, this difference 
appears  because  the poloidal magnetic field 
dominates the toroidal field near the light cylinder in the observer frame 
for the present case, 
and the poloidal and the toroidal fields 
are comparable to each other for the case in \citet{Ta06}. 
Although there is a large difference in emission direction near
 the ligh cylinder, the pulse profiles do not change
very much \citep{Ta06} because the radiation very close to the light 
cylinder mainly contribute to the bridge phase.  As we will 
see later, however, the calculated phase-resolved spectra  in the present case 
explain the observations  better than the 
results in \citet{Ta06}. Therefore, the present results suggest 
that the radiation diretion near the light cylinder seem
 to be in the azimuthal  direction  rather than in the radial direction. Recent 
particle simulation for the global structure of the charge separated 
magnetosphere done by \citet{WS07} 
also indicated such behavior of the emission direction. On the other hand, 
the time-dependent force free relativistic MHD solution obtained by 
\citep{Sp06} indicated the radial motion 
of the particles near the light 
cylinder. Furthermore Bucciantini et al. (2006)
have considered a more general relativistic MHD approach for rotating pulsars and their solutions asymptotically approach the force-free ones similar to that obtained by \citep{Sp06} in the high magnetized wind case, which is closed to our case. Therefore the global structure of the magnetic field is still an open question.

In this paper, we neglect the contribution from the inward emissions with the following reasons. In general, since the charged particles accelerate only within the gap and they lose energy during the cascade process, the incoming charged particles and hence the incoming photons cannot have an energy exceeding $\gamma _em_ec^2$ which is the energy of a charged particle immediately after it leaves the gap. $\gamma _e\sim 10^7$ is the local Lorentz factor of the charged particle upon leaving the outer gap (cf.~Equation~\ref{eq:gammae}). On the other hand, charged particles within the gap are being accelerated continuously and will gain an energy of $eV_{gap}$ with $V_{gap}\approx 6.6\times 10^{12}f_0^2B_{12}P^{-2}{\rm \;V}\sim 10^{15}$ V \citep{a29} where $f_0\approx 0.2$ is the average value of the local gap size at $R_L$. Therefore, we can roughly estimate the ratio of the intensity of the radiation due to the incoming photons to the radiation due to the outgoing photons by $(\dot{N}_{gap}\gamma _em_ec^2)/(\dot{N}_{gap}eV_{gap})\sim 0.5\%$. As a result, when we compute the light curve, we neglect the contribution from the incoming photons.

\begin{figure}
\epsscale{.8}
\plotone{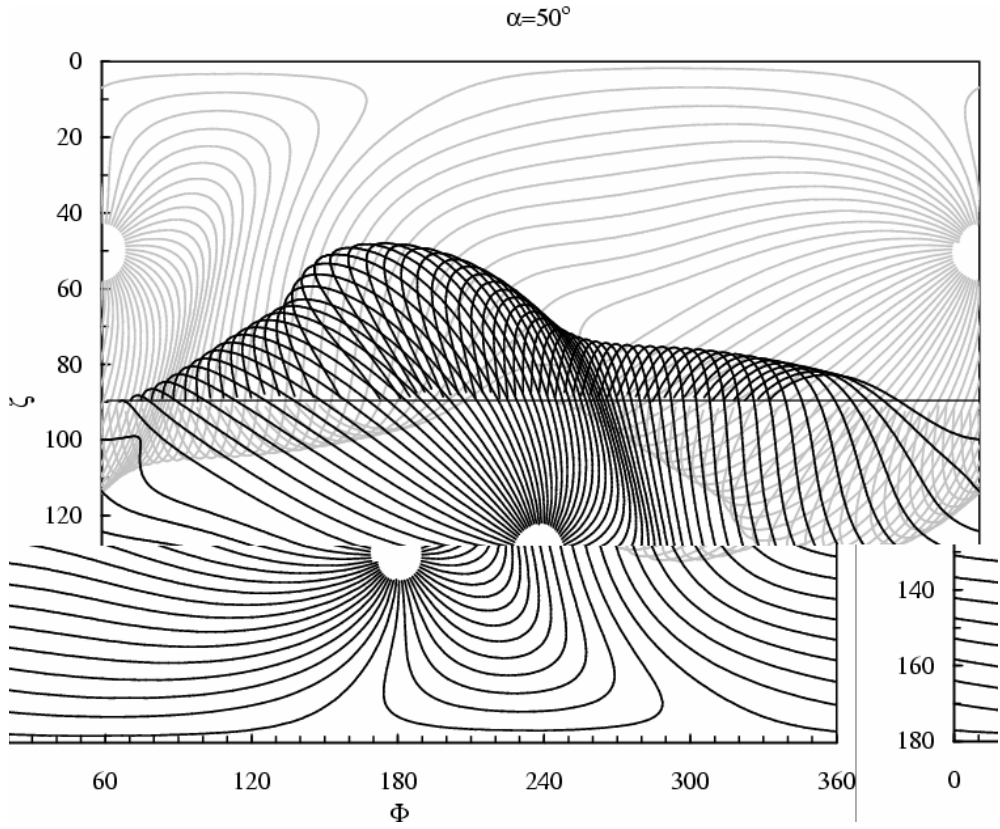}
\caption{Photon emission pattern for $\alpha=50^\circ$ and $a_1=0.97$.\label{fig:pattern}}
\end{figure}

By considering the radiation to be emitted tangent to the magnetic field lines in the co-rotating frame, we project the radiating points onto the $\zeta-\Phi$ plane. Figure~\ref{fig:pattern} shows the photon emission pattern for the inclination angle $\alpha=50^{\circ}$ and $a_1=0.97$. Here we assume the emission region to extend from the stellar surface to the light cylinder and assume a symmetry between the north and the south poles. In other words, when there is a photon emitting with $(\zeta,\Phi)$ from the north pole, there is another photon emitting with $(180^\circ-\zeta,180^\circ+\Phi)$ from the south pole as well. In Figure~\ref{fig:pattern}, the grey lines correspond to the outgoing photons emitted from the north pole, the pole which is making an acute angle with the rotational axis. The black lines correspond to the outgoing photons emitted from the south pole. For example, for an observer at a viewing angle smaller than $90^{\circ}$, the emission region by the north pole (grey) corresponds to the radiation emitted in the region between the inner boundary (stellar surface) and the null charge surface, and the emission region by the south pole (black) corresponds to the radiation emitted in the region beyond the null charge surface.

\begin{figure}
\epsscale{.8}
\plotone{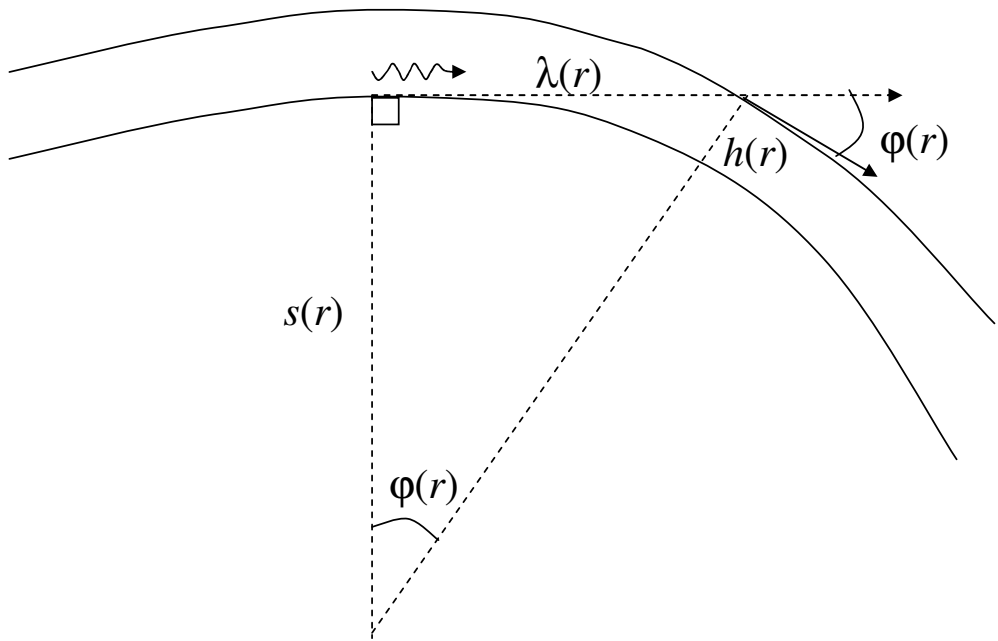}
\caption{Close-up illustration of two field lines that approximate two concentric circles.\label{fig:sin_beta}}
\end{figure}

The viewing angle which is the angle between the observer and the rotational axis can be set at a certain value $\zeta _0$. Then we can measure the number of photons traveling in this $\zeta _0$ direction and produce a theoretical light curve. However, before we move on to produce the theoretical light curve, we need to mention that the radiation is, in fact, emitted within a finite emission cone of half-angle $\varphi(r)$ instead of simply tangent to the field lines. The criteria for counting a photon becomes $\zeta-\varphi(r)\le\zeta _0\le\zeta+\varphi(r)$. This effect can be understood in a geometrical point of view. Figure~\ref{fig:sin_beta} shows a close-up illustration of two field lines that approximate two concentric circles. $h(r)$ is the local thickness of the outer gap, $s(r)$ is the local radius of curvature of the field line and $\lambda(r)$ is the pair creation mean free path. According to Figure~\ref{fig:sin_beta}, since the secondary pairs are produced just above the outer boundary of the outer gap accelerator, we can estimate the pitch angle of the new born pairs as
\begin{equation}
\sin^2\varphi(r)=\frac{2f(r)R_L}{s(r)}\label{eq:sinbeta2},
\end{equation}
where $f(r)$ is the fractional gap thickness defined by $f(r)=h(r)/R_{L}$. The self-sustained outer gap model of \citet{ZC97} estimated the fractional gap thickness as $f(R_{L}/2)\sim 5.5P^{26/21}B^{-4/7}_{12}$ which is $\sim 0.04$ for the Crab pulsar and $\sim 0.13$ for the Vela pulsar. By considering the conservation of magnetic flux along a field line, with 1-D approximation, $f(r)=f(R_L)\left(\frac{r}{R_L}\right)^{\frac{3}{2}}$ \citep{a1}. Since $s(r)=\sqrt{rR_L}$ in static dipole approximation, $\sin\varphi(r)=\sin\varphi(R_L)\left(\frac{r}{R_L}\right)^{1/2}$.

\subsection{Inner and outer boundaries
 of the outer gap\label{section:azimuth_width}}
\begin{figure}
\epsscale{1}
\plotone{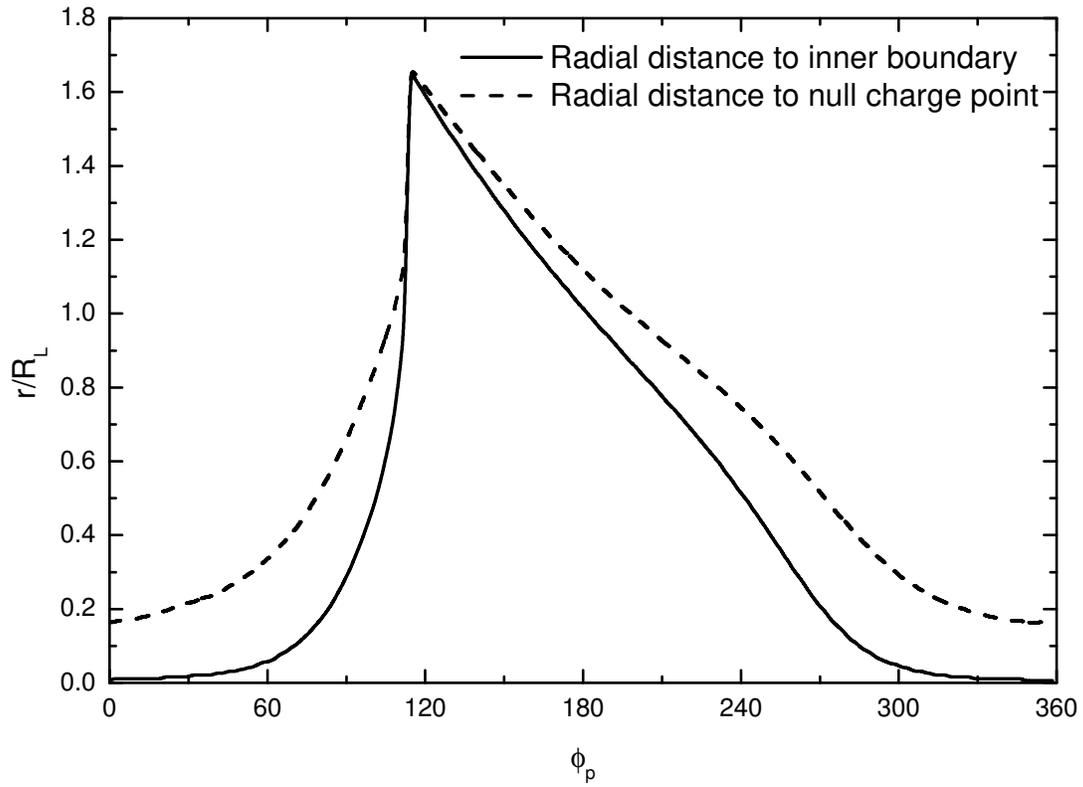}
\caption{The radial distance to the inner boundary and the null charge surface on the last closed field lines. $\phi _p$ refers to the polar cap angle. The angle $\phi _p=0^\circ$ represents the \mbox{\boldmath$\Omega$}-\mbox{\boldmath$\mu$} plane. \label{fig:inner}}
\end{figure}

As the electrodynamical studies have shown \citep{HHS03,Ta04,Ta06,b8}, the inner boundary of the outer gap accelerator is shifted inwards from the null charge surface with an increase in the current through the gap. In fact, if there is no current injection from the inner and the outer boundaries, the inner boundary will be located at a position on which $B_z/B=j_g$ is satisfied, where $j_g$ is the current density in unit of $\Omega B/2\pi$ carried by the pairs created in the gap and is constant along the field line for the steady state \citep{Ta04}. For example, if no current is created in the gap ($j_g=0$), the inner boundary is located at the position where $B_z/B=0$ is satisfied, that is, on the null charge surface. On the other hand, if $j_g\sim\cos\alpha$ on a particular magnetic field line, the inner boundary on the field line is located at the stellar surface, where $B_z/B\sim\cos\alpha$ is satisfied. We expect that the created current density is proportional to the pair-creation rate, which depends on the radial distance as $r^{-3/8}$ \citep{a1}. Since most of the pairs are created around the null charge surface in the outer gap, we may be able to relate the created current density with the radial distance to the null charge surface of each last closed field line as $j_g(\phi _p)=j_g(0)[r_n(0)/r_n(\phi _p)]^{-3/8}$, where $j_g(0)$ and $r_n(0)$ are, respectively, the created current density on and the radial distance to the null charge surface on the last closed field line with the polar cap angle $\phi _p=0^\circ$. We know the location of the inner boundary with the azimuthal angle if we would estimate the created current density $j_g(0)$. As demonstrated by the electrodynamical studies, however, the current structure in the gap is very sensitive to the gap geometry such as trans-field thickness and the longitudinal width. Since the gap geometry in the pulsar magnetosphere should be determined by the global condition \citep{WS07} and since there is no study for the 3-dimensional magnetosphere of an inclined rotator, we deal with the created current $j_g(0)$ by using some model parameters. Figure~\ref{fig:inner} summarizes the variation of the radial distances to the inner boundary and the null charge surface on the last closed field lines against the polar cap angle of the field lines around the magnetic axis. From Figure~\ref{fig:inner}, we can see that only a small current is created around $\phi _p\sim 180^{\circ}$, so the outer gap must be less active there. In this paper, we constrain the width of the polar cap angle with the field lines on which the pair creation mean-free path, $\lambda(r)\sim [2s(r)f(r)R_L]^{1/2}\sim 2 f^{1/2}(R_L/2)r$, at the null charge surface of the active field lines is estimated to be shorter than $R_L$. This condition produces an azimuthal extension of the outer gap of $\Delta\phi _p\sim 250^{\circ}$.

For the outer boundary, the position should be determined with the global 
model such as \citet{WS07} with current. For the present local 
model, the position is a free parameter and  
we put it at the light cylinder because we assume that the emissivity of the curvature 
radiation is declined rapidly near/beyond the light cylinder and/or the radiations beyond 
the light cylinder are beamed 
out of  line of sight due to the  magnetic bending. We may put the outer  
boundary inside of the light cylinder.
 However, the resultant  pulse profile and the spectra does 
not change very much unless the outer boundary is located close 
to the null charge surface, because 
the accelerated particles inside of the gap contribute to the 
total radiation emission outside the outer gap \citep{WS07}. 

\subsection{Light curve}
\begin{figure}
\epsscale{1}
\plotone{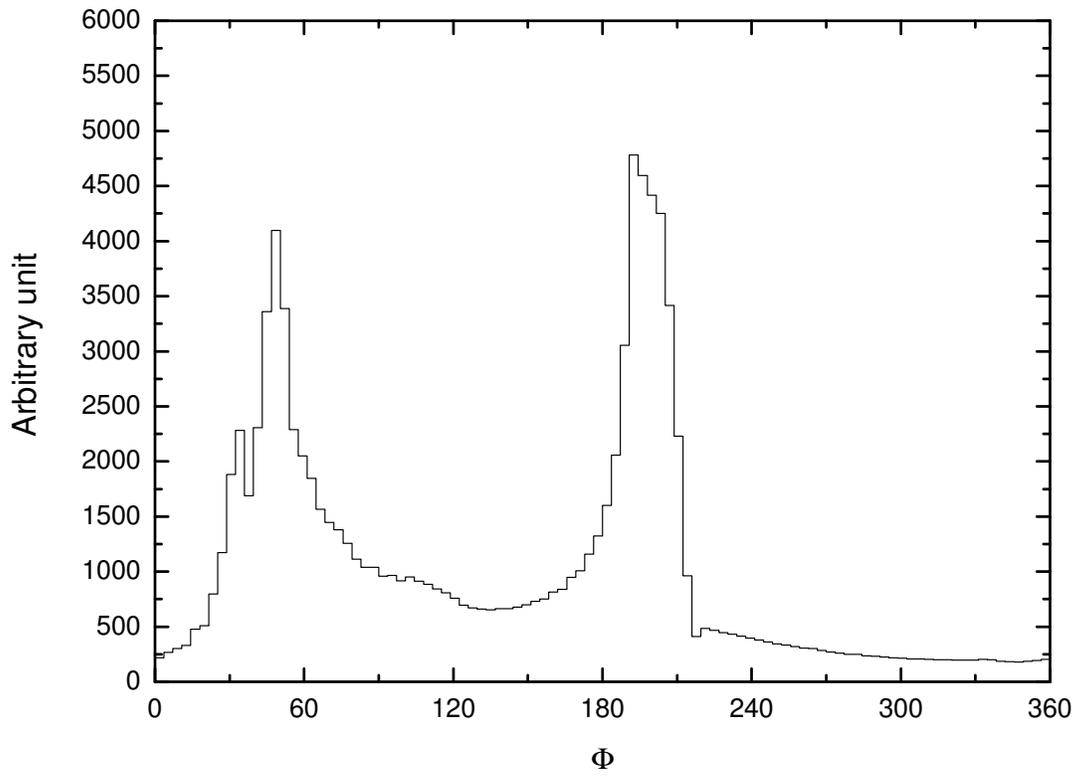}
\caption{Theoretical light curve for the Crab pulsar. The fitting parameters are $\alpha=50^\circ$, $a_1=0.97$, $\zeta _0=76^\circ$ and $\sin\varphi(R_L)=0.04$. \label{fig:lightcurve}}
\end{figure}

\begin{figure}
\epsscale{1}
\plotone{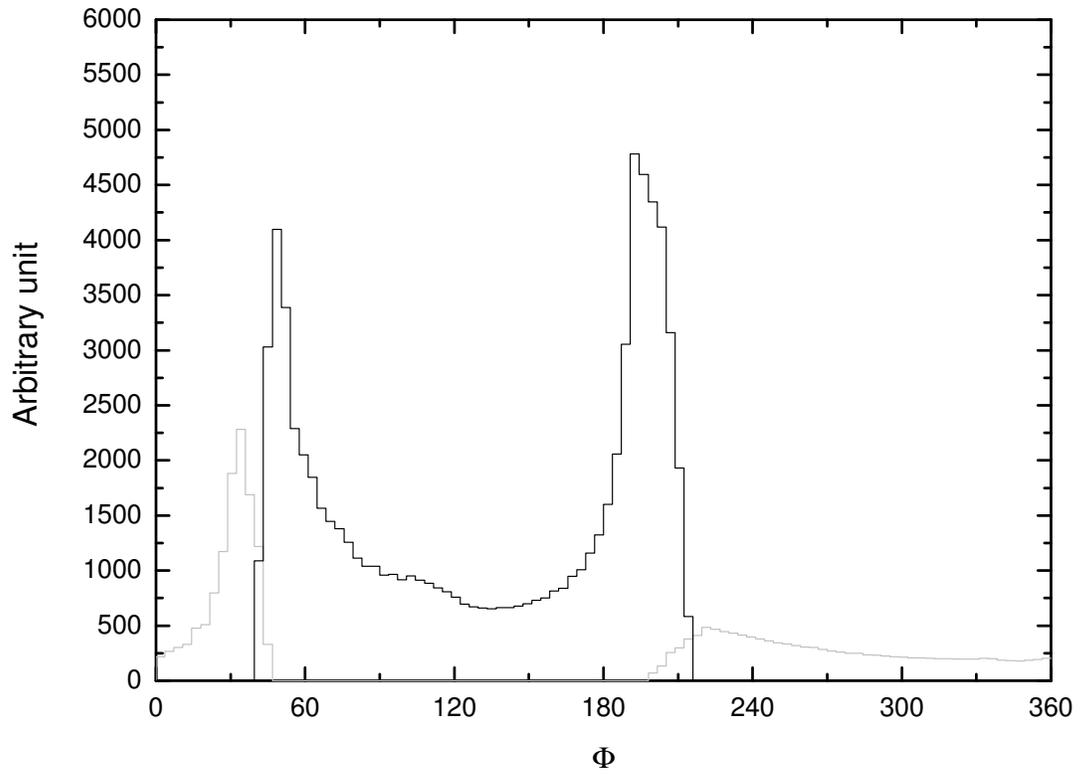}
\caption{The breakdown of the light curve in Figure~\ref{fig:lightcurve} to show the contribution from each pole.\label{fig:lightcurvepoles}}
\end{figure}

Since the general feature of the light curve such as the standing phase of the pulse is mainly affected by the geometry of the emission regions, we produce a theoretical light curve by assuming constant emissivity. Figure~\ref{fig:lightcurve} shows the theoretical light curve for a pulsar with the inclination angle $\alpha=50^\circ$, $a_1=0.97$, viewing angle $\zeta _0=76^\circ$ and the azimuthal extension of the outer gap $\Delta\phi _p=250^\circ$. The pitch angle at $R_L$ is treated as a fitting parameter and is assumed to be $\sin\varphi(R_L)=0.04$. These parameters are chosen so that the modeled light curve explains the general features of the observation such as two peaks in a single period with a phase separation of $\sim 140^\circ$ between the two peaks. The breakdown of the light curve to show the contribution from the two poles separately is given in Figure~\ref{fig:lightcurvepoles}. The color scheme is the same as the one for the emission pattern in Figure~\ref{fig:pattern}. As we shall see later, the inclination angle and the viewing angle chosen to explain the observed light curve also produces phase-resolved spectra which are consistent with observation.

\section{Energy spectra of the observed photons \label{section:spectra}}
\subsection{Acceleration and emission in the gap}
We adopt the local electric field equation in the CHR model for the region beyond the null charge surface. By assuming that the local electric field is decreasing in a quadratic form for the region between the null charge surface $r_{\rm null}$ and the inner boundary of pair production region $r_{\rm in}$, we have
\begin{equation}E_\parallel=\left\{
\begin{array}{ll}
\frac{\Omega B(r)h^2(r)}{cs(r)} & r\ge r_{\rm null},\\
E_\parallel(r_{\rm null})
\frac{\left[\left(\frac{r}{r_{\rm in}}\right)^2-1\right]}
{\left[\left(\frac{r_{\rm null}}{r_{\rm in}}\right)^2-1\right]} & r<r_{\rm null}.
\end{array}\right.
\label{eq:Er}
\end{equation}
For $r\ge r_{\rm null}$, $E_\parallel$ is the vacuum solution given in \citet{CHRa}. The vacuum solution is a good approximation for pulsars with thin gaps like the Crab pulsar.

The radial distance $r$ to the null charge surface varies with the field lines and the local curvature radius also depends on the field lines. Therefore, strictly speaking, the electric field component along a magnetic field is a function of both the radial distance $r$ and the polar cap angle $\phi _p$, that is, $E_\parallel=E_\parallel(r,\phi _p)$.

\begin{figure}
\epsscale{1}
\plotone{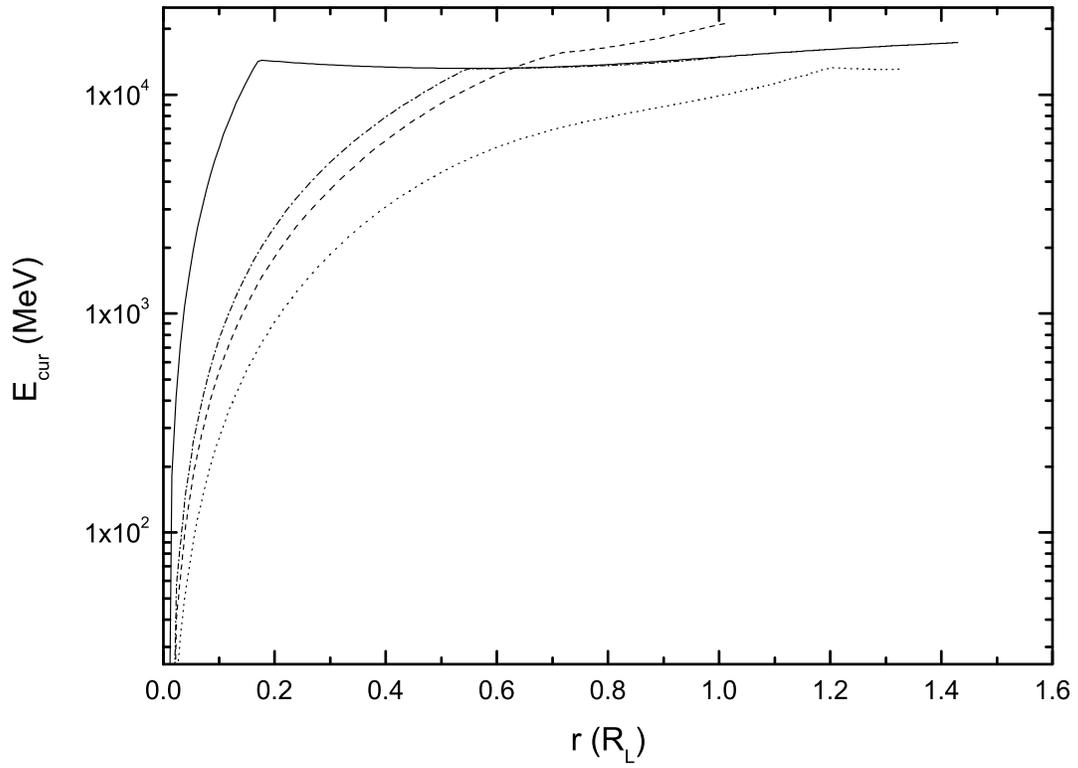}
\caption{Curvature photon energy vs $r$ for the field lines at $\phi _p=0^\circ$, $90^\circ$, $180^\circ$ and $270^\circ$ represented by solid, dashed, dotted and dash-dotted lines, respectively.\label{fig:E_cur}}
\end{figure}

When a relativistic charged particle which is accelerated continuously along a magnetic field line by the strong local electric field radiates by means of curvature radiation, the power gained by the accelerated particle as it goes through the electric potential, $eE_\parallel(r,\phi _p)c$, is transformed to the power radiated as curvature radiation in order to maintain the equilibrium. The total radiated power for each particle is $l_{cur}(r)=2 e^2c \gamma _e^4(r)/3s^2(r)$. The local Lorentz factor $\gamma _e$ of the primary particles can be found by requiring $eE_\parallel(r,\phi _p)c=l_{cur}$, and hence,
\begin{equation}
\gamma _e(r)=\left[\frac{3}{2}\frac{s^2(r)}{e}E_\parallel(r,\phi _p)\right]
^{1/4}.\label{eq:gammae}
\end{equation}
The characteristic energy of the radiated curvature photons is given by $E_{cur}(r)=\frac{3}{2}\hbar c\frac{\gamma^3_e(r)}{s(r)}$ and Figure~\ref{fig:E_cur} shows the variation of $E_{cur}(r)$ along the field lines with the polar cap angle of $\phi _p=0^{\circ}$ (solid-line), $90^{\circ}$ (dashed-line), $180^{\circ}$ (dotted-line) and $270^{\circ}$ (dashed-dotted line). From Figure~\ref{fig:E_cur}, we find that the particles are accelerated up to the ultra-relativistic regime so that $10$~GeV photons are emitted in the outer gap accelerator of the Crab pulsar via curvature process.

\subsection{Synchrotron radiation and inverse Compton scattering from secondary pairs}
Since the relation $E_{cur}(r) dN_\gamma/dt\sim l_{cur}(r)N$ is satisfied, the radiation spectrum of the primary particles in a unit volume is approximately described by
\begin{equation}
\frac{d}{dV}\left(\frac{d^2N_\gamma}{dE_\gamma dt}\right)\approx\frac{l_{cur}(r)n}{E_{cur}(r)}\frac{1}{E_\gamma}\label{eq:d2NdVdEdt}
\end{equation}
where $n=dN/dV$ is the number density of the primary particles. Here, we use $n=\Omega B(r)/2\pi ec$ which comes from the local Goldreich-Julian number density disregarding the angle between the local magnetic field direction and the rotational axis. Since the energy of each photon comes from curvature radiation, $E_\gamma\le E_{cur}(r)$. These primary curvature photons collide with the soft photons produced by synchrotron radiation of the secondary e$^+$ or e$^-$ to produce even more secondary e$^\pm$ pairs. In this way, the synchrotron photons become abundant and nearly all the curvature photons are converted into secondary e$^\pm$ pairs. As a result, the energy distribution of the secondary e$^+$ or e$^-$ is
\begin{eqnarray}
\frac{dN(r)}{dE_e}&\approx&\frac{1}{\dot{E}_e}\int _{E_e}^{E_{max}(r)}\frac{d^2N_\gamma(E_\gamma=2E'_e)}{dE_\gamma dt}(1-e^{-\tau _{\gamma\gamma}(E_\gamma,r)})dE'_e \nonumber \\
&\approx&\frac{1}{\dot{E}_e}\frac{l_{cur}(r)\Omega B(r)}{2\pi ecE_{cur}(r)}
\Delta V(r)\ln\left(\frac{E_{cur}(r)}{E_e}\right)\ {\rm for}\ \tau _{\gamma\gamma}\rightarrow\infty.\label{eq:dNdEe}
\end{eqnarray}
where the upper integration limit $E_{max}(r)$ is taken to be $E_{cur}(r)/2$ as the energy of each photon is divided between an e$^\pm$ pair, $E_e$ represents the energy of each e$^+$ or e$^-$ and $\dot{E}_e=2e^4B^2(r)\sin^2\beta(r) E_e^2 /3m^4_ec^7$ is the synchrotron energy loss rate. $\tau _{\gamma\gamma}(E_\gamma,r)$ is the attenuation depth for the absorption of the curvature photons. It will be very large in most of the magnetospheric regions except in the phases Trailing Wing 1, Bridge and Leading Wing 2. Detailed discussion on the absorption of curvature photons will be given in Section~\ref{section:absorb_cur}.

In the computer program, we divide the polar cap into $N_B$ equal divisions so that each division spans an angle of $\Delta\phi _p=360^\circ/N_B$. Therefore, each division, represented by a magnetic field line, will take up a magnetic flux of $\Phi _{Gap}/N_B$. The volume element $\Delta V(r)$ is represented by $\Delta A(r)\Delta l(r)$ where $\Delta A(r)$ and $\Delta l(r)$ mean the area and the length of the tube-like volume element along a field line where observable photons are produced. Due to magnetic flux conservation, $B(r)\Delta A(r)=\Phi _{Gap}/N_B$ such that Equation~(\ref{eq:dNdEe}) becomes
\begin{equation}
\left(\frac{dN(r)}{dE_e}\right)_i\sim\frac{1}{\dot{E}_e}\frac{l_{cur}(r)\Omega \Phi _{Gap}\Delta l(r)}{2\pi ecE_{cur}(r)N_B}\ln\left(\frac{E_{cur}(r)}{E_e}\right)
\end{equation}
where $\Phi _{Gap}=2\pi f(R_L)B(R_L)R_L^2$.

The photon spectrum of the synchrotron radiation is given by
\begin{eqnarray}
F_{syn}(E_\gamma ,r)&=&\sum _{i=1}^{N_B}\int _{E_{min}}^{E_{max}}\left(\frac{dN(r)}{dE_e}\right)_i\left[\frac{d^2N_\gamma}{dE_\gamma dt}\right]_{syn}dE_e\nonumber\\
&=&\frac{\sqrt{3}e^3B(r)\sin\varphi(r)}{m_ec^2h}\frac{1}{E_\gamma}\sum _{i=1}^{N_B}\int _{E_{min}}^{E_{max}}\left(\frac{dN(r)}{dE_e}\right)_iF(x)dE_e
\end{eqnarray}
where $F(x)=x\int _x^\infty K_{\frac{5}{3}}(\xi)d\xi$, $K_{\frac{5}{3}}$ is the modified Bessel function of order $5/3$, $x=E_\gamma/E_{syn}(r)$ and
\begin{equation}
E_{syn}(r)=3\hbar eB(r)\sin\varphi(r)E_e^2/2m_e^3c^5\label{eq:Esyn}
\end{equation}
is the critical synchrotron photon energy. For the integration, the upper integration limit $E_{max}=E_{max}(r)=E_{cur}(r)/2$, while the lower integration limit is chosen in a way that $E_{min}/m_ec^2=20$ \citep{Ta06}.

Similarly, the photon spectrum due to inverse Compton scattering is given by
\begin{equation}
F_{ICS}(E_\gamma ,r)=\sum _{i=1}^{N_B}\int _{E_{min}}^{E_{max}}\left(\frac{dN(r)}{dE_e}\right)_i\left[\frac{d^2N_\gamma}{dE_\gamma dt}\right]_{ICS}dE_e.\label{eq:FICS}
\end{equation}
If $E_e/m_ec^2>>1$, the spectrum of the inverse Compton scattered photons per electron (cf.~\S 2.7 in \citeauthor{a19}, \citeyear{a19}) is
\begin{eqnarray}
\left[\frac{d^2N_\gamma}{dE_\gamma dt}\right]_{ICS}&=&\int _{\epsilon _1}^{\epsilon _2}\frac{3\sigma _Tc}{4(E_e/m_ec^2)^2}\frac{[n_{syn}(\epsilon ,r)+n_X(\epsilon ,r)]}{\epsilon}d\epsilon\nonumber\\
&&\times\left[2q\ln q+(1+2q)(1-q)+\frac{(\Gamma q)^2(1-q)}{2(1+\Gamma q)}\right]\label{eq:qq}
\end{eqnarray}
where $q=E_1/\Gamma(1-E_1)$, $\Gamma=4(\epsilon/m_ec^2)(E_e/m_ec^2)$ and $E_1=E_\gamma/E_e$. There are two possible soft photon sources for the inverse Compton scattering. The thermal photons with typical temperature $T$ produced by the stellar surface produce a dominant absorption effect for photons emitted from locations close to the star. The number density of this kind of photons is given by
\begin{equation}
n_{X}(\varepsilon,r)=\frac{1}{\pi^2(\hbar c)^3}\frac{\varepsilon^2}{\exp(\varepsilon/kT)-1}\left(\frac{R}{r}\right)^2.
\end{equation}
The other source of soft photons arises from synchrotron radiation. The absorption due to this kind of photons is more significant near the light cylinder. The number density of the synchrotron photons, $n_{syn}(\epsilon ,r)$ is described by
\begin{equation}
n_{syn}(\epsilon ,r)=\frac{F_{syn}(\epsilon ,r)}
{cr^2\Delta\Omega(r)},\label{eq:nsyn}
\end{equation}
where $\Delta\Omega(r)$ is the solid angle of the beam of synchrotron photons and is estimated as
\begin{equation}
\Delta\Omega(r)=\int _0^{2\pi}\int _0^{\varphi(r)}d\phi\sin\theta d\theta
\approx\pi\varphi^2(r).\label{eq:dWr}
\end{equation}
$\dot{E}_e$ in $\frac{dN(r)}{dE_e}$ in Eq.~(\ref{eq:FICS}) is the sum of both the synchrotron energy loss rate and the inverse Compton energy loss rate due to thermal photons given by $\dot{E}_e=\frac{\sigma _T(m_eckT)^2}{16\hbar^3}(\ln(4\gamma _ekT/m_ec^2)-5/6-0.5772-0.5700)$ \citep{a19}. The surface temperature of the Crab pulsar is taken to be  $2\times 10^6$~K \citep{TBJEK}.

In order that Equation~(\ref{eq:qq}) is valid, $q$ must be positive, so $E_1<1$. Moreover, since the number of photons cannot be negative, we require the value in the bracket of Equation~(\ref{eq:qq}) to be greater than or equal to zero. The upper integration limit $\epsilon _2$ is chosen in such a way that $F_{syn}(\epsilon _2,r)$ is very small for that particular field line; whereas the lower integration limit $\epsilon _1$ is taken to be $>1$ eV.

\subsection{Absorption of curvature photons\label{section:absorb_cur}}
 In our previous studies (Cheng, Ruderman \& Zhang 2000), 
we have anticipated that most of 
the curvature photons will be converted into the secondary pairs 
through the pair-creation 
process with the magnetospheric X-rays. The typical pair-creation 
mean free path, which is estimated from $l^{-1}(R_{lc}/2)
\sim (1-\cos\theta_{col}) n_X(R_{lc}/2)\sigma_{\gamma\gamma}$,
 where $n_X(R_{lc}/2)$ is the typical non-thermal X-ray number density,  
 $\sigma_{\gamma\gamma}\sim \sigma_{T}/3$ is the pair-creation cross section 
with $\sigma_{T}$ being Thomson cross section, and $\theta_{col}(R_{lc}/2)
\sim \sqrt{2f(R_{lc}/2)R_{lc}/s(R_{lc}/2)}\sim \sqrt{0.2}$ 
is the typical collision angle between 
the magnetospheric X-ray and the $\gamma$-rays emitted in the gap. 
For the Crab pulsar, the typical number density of X-rays is 
$n_{X}\sim L_{X}(<E_X>)/\delta\Omega (R_{lc}/2)^2
c <E_X>\sim 6\times10^{17}~\mathrm{cm}^3$, where  we used the typical 
energy $<E_X>\sim (2m_ec^2)^2/(1-\cos\theta_{col})/10
~\mathrm{GeV}\sim260$~eV, the typical non-thermal X-ray luminosity $L_X\sim
5\times10^{34}\mathrm{erg/s}$, and the solid angle $\delta\Omega=1$~radian.
 As a result, the mean free path  becomes $l\sim 2\times
10^{7}~\mathrm{cm}$ ($\sim R_{lc}/5$) for the 10~GeV photons,
 and therefore we have believed that most of the
 10~GeV photons emitted are converted  
into the secondary  pairs. In the present paper, on the other hand, 
 we take into the fact that some
curvature photons emitted far away from the star (or near the light cylinder) 
could avoid the photon-photon pair creation process due to shorter escape distance and
lower photon density near the light cylinder. In the present paper, we explicily calculate the optical depth for photons emitted at a given position. We can show that indeed some 10~GeV photons emitted near the light cylinder and
almost all photons with energy below GeV can  escape from the magnetosphere because their mean-free path becomes longer than the light radius. These survival curvature photons will contribute to the high energy peaks around $10$ GeV in the Trailing Wing 1, Bridge and Leading Wing 2 phases. Following \citet{DC97}, the photon spectrum of the surviving $\gamma$-ray photons is obtained by 

\begin{equation}
F_{cur,sur}=F_{cur}e^{-\tau(E_\gamma,r)}
\end{equation}
where $F_{cur}$ is the curvature spectrum and is described by
\begin{equation}
F_{cur}(E_\gamma,r)=\frac{dN\sqrt{3}e^2\gamma _e}{2\pi\hbar sE_\gamma}F(x)
\end{equation}
where $dN$ is the number of primary e$^\pm$ pairs in the emission region and is given by $dN=n_{\rm{GJ}}\Delta A\Delta l$ and $x=E_\gamma/E_{cur}(r)$. The attenuation depth $\tau(E_\gamma,r)$ is calculated from \citep{JR76}
\begin{equation}
\tau(E_\gamma,r)=l(r)\int ^{\varepsilon _{max}}_{\varepsilon _{min}}
[n_{syn}(\varepsilon,r)+n_X(\varepsilon,r)]\sigma _{\gamma\gamma}(E_\gamma,\varepsilon)d\varepsilon
\end{equation}
with $l(r)$ being the distance between the emission location and the light cylinder and $\sigma _{\gamma\gamma}(E_\gamma,\varepsilon)$ being the cross section for photon-photon pair creation and is given by
\begin{equation}
\sigma _{\gamma\gamma}(E_\gamma,\varepsilon)=\frac{3}{16}\sigma _T(1-v^2)\left[(3-v^4)\ln\left(\frac{1+v}{1-v}\right)-2v(2-v^2)\right]
\end{equation}
where
\begin{equation}
v=\sqrt{1-\frac{m_ec^2}{E_\gamma\varepsilon}}.
\end{equation}

The total photon flux received on Earth is
\begin{equation}
F(E_\gamma)=\frac{1}{\bigtriangleup\Omega D^2}\sum _r\left[F_{syn}(E_\gamma,r)+F_{ICS}(E_\gamma,r)+F_{cur,sur}(E_\gamma,r)\right]\label{eq:F
_Egamma}
\end{equation}
where $D=2$ kpc is the distance of the Crab pulsar from the Earth and $\bigtriangleup\Omega$ is the solid angle chosen to be $1$ sr for simplicity.

\subsection{Phase-resolved spectra}
The computation of the photon flux is divided into four steps. Firstly, we calculate the synchrotron radiation photon flux. Then, we use this synchrotron flux to calculate the inverse Compton scattering photon flux. Next, we adjust the peak intensities of the synchrotron and the inverse Compton scattering spectra by requiring $\int E_\gamma F(E_\gamma)dE_\gamma$ for synchrotron spectrum alone before the consideration of the inverse Compton scattering radiation to be the same as the sum of the synchrotron and the inverse Compton scattering spectra. However, the relative peak intensities of the synchrotron and the inverse Compton scattering spectra are kept unchanged. We employ this third step to comply with energy conservation because all the energy in inverse Compton scattered photons comes from the synchrotron photons. Finally, we calculate the survival curvature radiation.

Figure~\ref{fig:spec_averaged1} shows the observed data of the phase-averaged spectrum of the Crab pulsar and the theoretical fitting spectrum with $f(R_{L})=0.2$ and  $\sin\varphi(R_{L})=0.06$, calculated by using the synchrotron self-Compton mechanism together with the survival curvature photons from $100$~eV to $10$~GeV. The blue and red lines show the emission beyond the null charge surface (i.e.~from the south pole) and between the inner boundary and the null charge surface (i.e.~from the north pole), respectively. For each colour, the dashed line represents the synchrotron spectrum, the dotted line represents the inverse Compton scattering spectrum, the dash-dotted line represents the survival curvature spectrum and the black solid line is the sum of the three spectra from both poles. The fitting spectrum explains the observation from $100$~eV to $10$~GeV.

\begin{figure}
\epsscale{.8}
\plotone{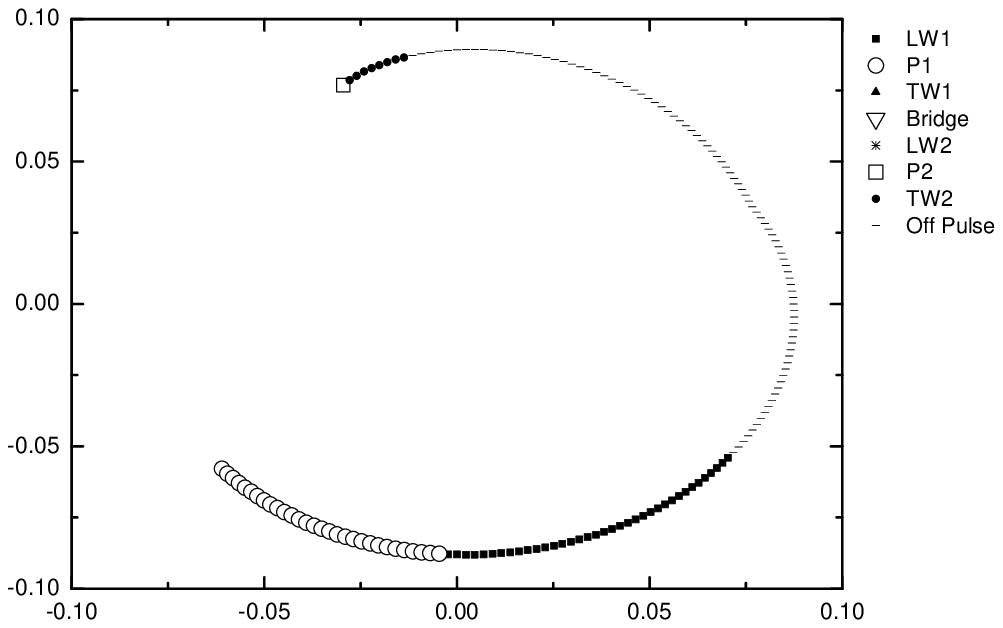}
\plotone{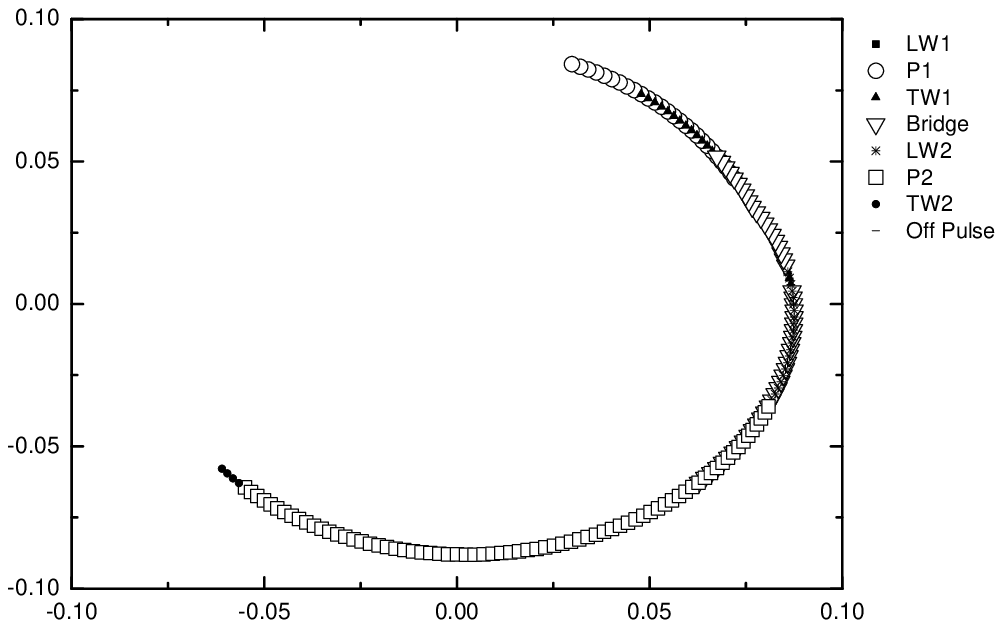}
\caption{Footprints that correspond to particular phases. The upper panel is the north pole and the lower panel is the south pole. $\alpha=50^\circ$, $a_1=0.97$ and $\varsigma=76^\circ$. Some footprints do not belong to any phase and some of them belong to more than one phases. From $\phi _p=73.5^\circ$ to $223.5^\circ$, there is no radiation produced along the line of sight. \label{fig:NSPoles}}
\end{figure}

Figure~\ref{fig:spectra_consistent} shows the predicted phase-resolved spectra as a break-down of the phase-averaged spectrum in Figure~\ref{fig:spec_averaged1}. The colour and line schemes are the same as in Figure~\ref{fig:spec_averaged1}. The phase intervals are defined in the same way as in \citet{FMNT98}. The pulse of the Crab pulsar is divided into 8 phases, namely Leading Wing 1, Peak 1, Trailing Wing 1, Bridge, Leading Wing 2, Peak 2, Trailing Wing 2 and Off Pulse. The criteria for the division is listed in Table~\ref{table:Kuiper}. We choose the azimuthal angle of $52^\circ$ from the $\mbox{\boldmath$\Omega$}$-$\mbox{\boldmath$\mu$}$ plane to be phase zero. The footprints corresponding to each phase are plotted for both poles in Figure~\ref{fig:NSPoles}. Although our prediction is not too good (for example, in LW1, the peak at about $100$~MeV cannot be produced;
 whereas in P2 and TW2, the peaks at about $100$~MeV are much too high), most of the spectral features can be explained. However, it has been argued that the north pole and the south pole need not to be perfectly symmetric. Since the dipole may not be at the centre of the star, just like the sunspot geometry \citep{ruderman91}, it is possible that the two poles could have small differences. Furthermore, $\sin\varphi(R_L)$ is calculated according to a simple static dipolar field, which is clearly not a good approximation in the outer magnetosphere of a realistic rotating dipolar field with current flow. In the best fit, we thus allow the gap sizes to be different for the two poles and the pitch angles to vary for individual phases.

Figure~\ref{fig:spectra} shows the comparison between the observed data of the phase-resolved spectra of the Crab pulsar and the theoretical best fitting results. Table~\ref{table:fitting} summarizes the values of $f(R_L)$ and $\sin\varphi(R_L)$ for different phases for each pole.  We find that the fitting pitch angles for the phases consisting of field lines at a polar angle closer to $\phi _p=0^\circ$ are smaller, while those for the phases consisting of field lines at a polar angle closer to $\phi _p=180^\circ$ are larger. This agrees with the theory that the outer gap is thinner around $\phi _p=0^\circ$ and thicker around $\phi _p=180^\circ$.

As we consider the emission region to extend inwards to an inner boundary inside the null charge surface, we can fit the spectrum of LW1. This cannot be done in \citet{a1} which considered the emission region to be from null charge surface to light cylinder alone, i.e.~one pole only. Figure~\ref{fig:spectra} shows that this phase is mainly contributed by the radiation from the north pole. Another phase that cannot be produced in \citet{a1} is TW2. \citet{a1} estimated $\Delta\phi _p\sim 160^\circ$, but in Section~\ref{section:azimuth_width}, we estimated that $\Delta\phi _p$ can be $\sim 250^\circ$ for the extended new emission geometry. The widening of the azimuthal extension of the outer gap allows us to obtain a reasonable fitting for the phase TW2. 

\begin{figure}
\epsscale{1}
\plotone{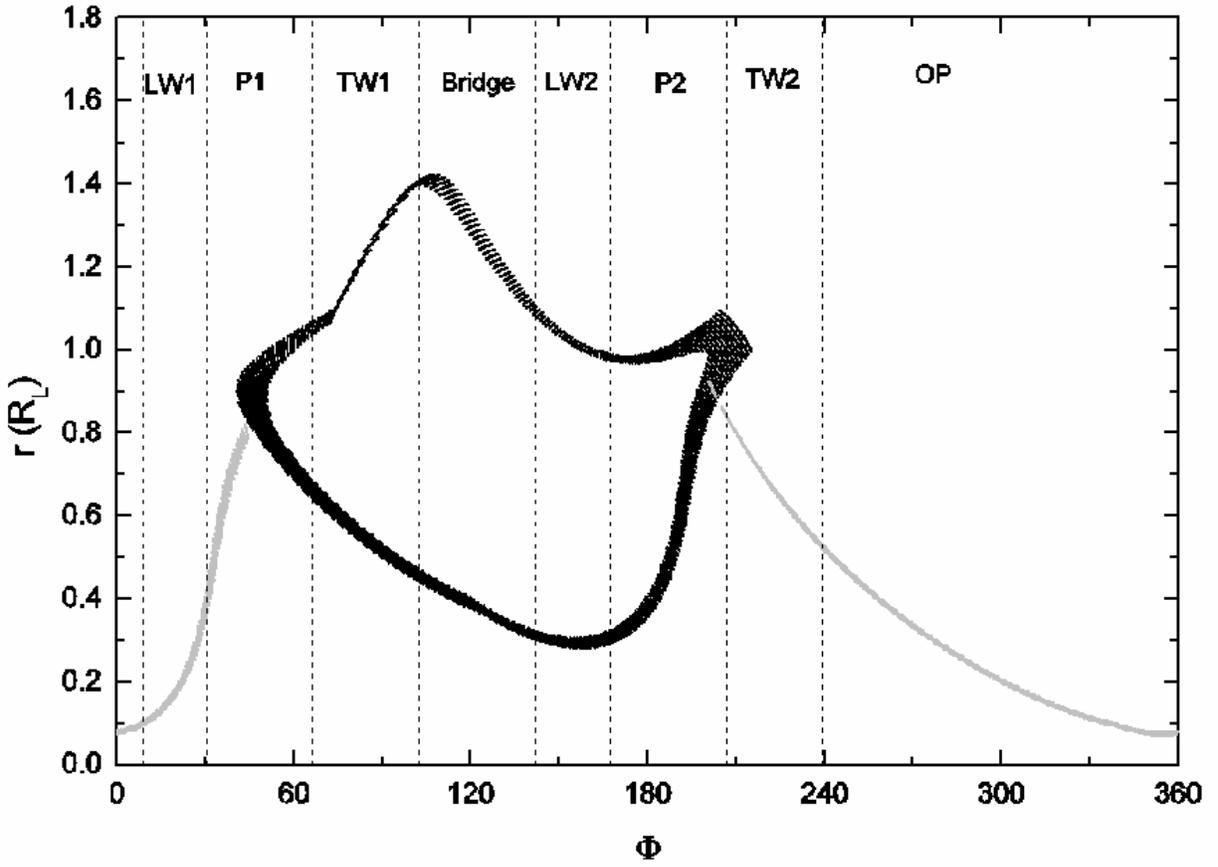}
\caption{The emission locations for photons which have been counted in Figure~\ref{fig:lightcurvepoles}.\label{fig:phi_r}}
\end{figure}

\begin{figure}
\epsscale{.80}
\plotone{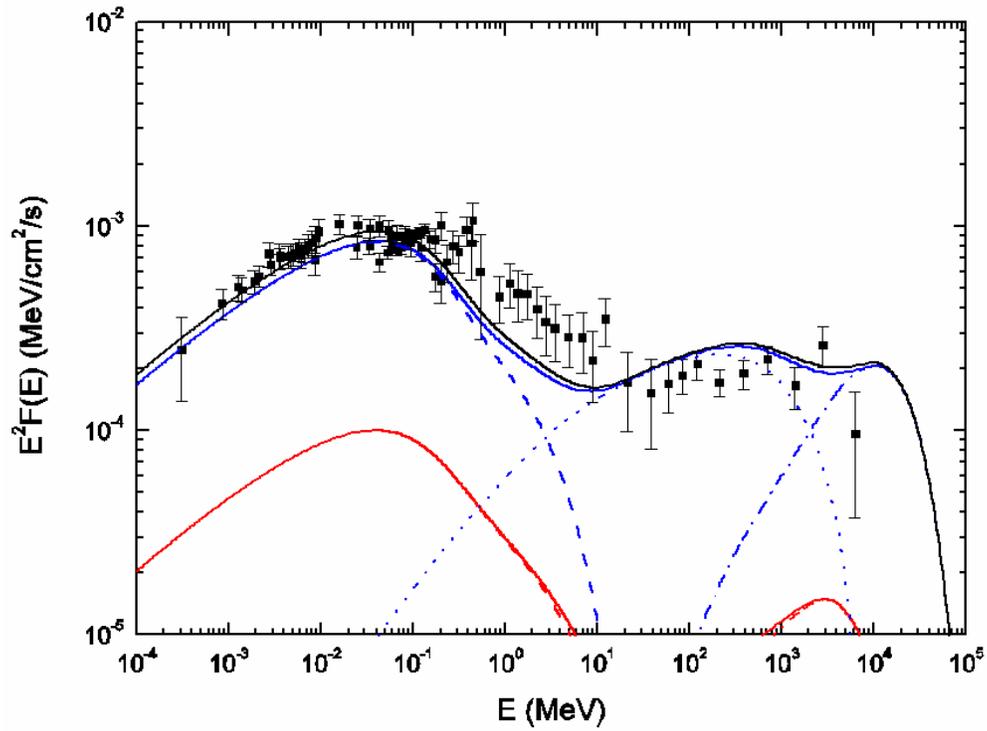}
\caption{Phase-averaged spectrum of the Crab pulsar. The observed data are taken from \citet{Ku01}. The fitting parameters are $\alpha=50^\circ$, $a_1=0.97$, $\varsigma=76^\circ$, $f(R_L)=0.2$ and  $\sin\varphi(R_L)=0.06$. \label{fig:spec_averaged1}}
\end{figure}

\begin{figure}
\includegraphics[angle=90,scale=1.8]{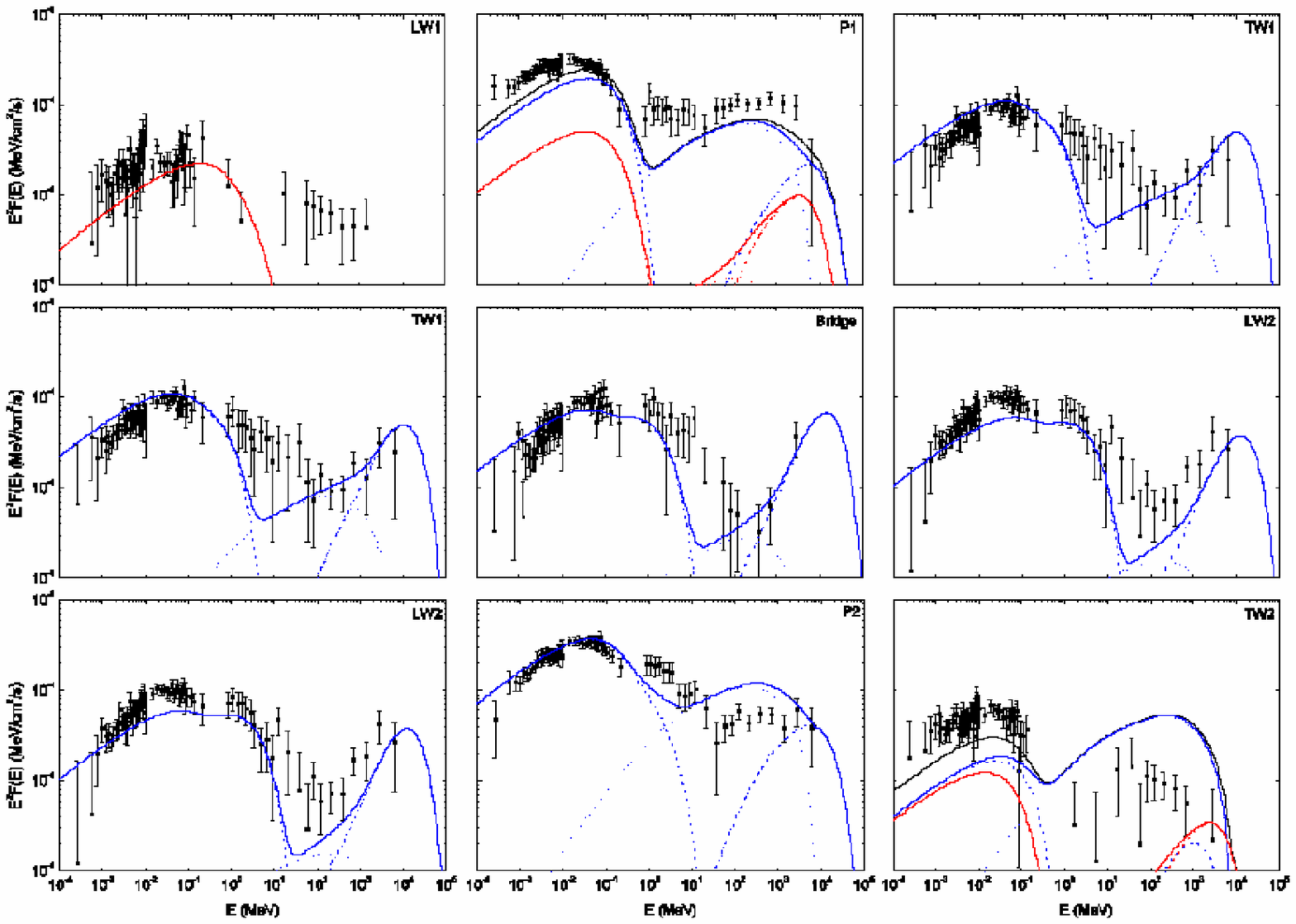}
\caption{The phase-resolved spectra with $\alpha=50^\circ$, $a_1=0.97$, $\varsigma=76^\circ$, $f(R_L)=0.2$ and  $\sin\varphi(R_L)=0.06$ as a breakdown of Figure~\ref{fig:spec_averaged1}. \label{fig:spectra_consistent}}
\end{figure}

From Figure~\ref{fig:spectra}, we can see that except LW1 and the $100$ MeV regime of P1, all the phase-resolved spectra are dominated by the radiation from the south pole. In general, the spectra are mainly contributed by the part from the null charge surface to the light cylinder, so the results in \citet{a1} are good approximations. The peaks at the high energy regime near several GeV in TW1, Bridge and LW2 are believed to be the survival curvature radiation emitted at locations far away from the star (cf.~Figure~\ref{fig:phi_r}). This agrees with the model that most of the curvature radiation has been converted into e$^\pm$ pairs and then emitted as synchrotron and inverse Compton scattering photons.

For the 3 phases between the peaks, i.e.~in the middle row of Figure~\ref{fig:spectra}, namely TW1, Bridge and LW2, the synchrotron peaks from the south pole are much wider than their counterparts from the north pole. In fact, they show a kind of double-peak feature with a plateau around $1$ MeV. According to Figure~\ref{fig:phi_r}, the radiation of each of these 3 phases are actually emitted from two well-separated regions. For TW1, the radiation is emitted from an outer region around $1.15R_L<r<1.7R_L$ and an inner region around $0.46R_L<r<0.67R_L$. For Bridge, they are $1.28R_L<r<1.72R_L$ and $0.31R_L<r<0.45R_L$. For LW2, they are $1.15R_L<r<1.27R_L$ and $0.29R_L<r<0.31R_L$. In order to know more clearly this feature, we plot a further decomposed phase-resolved spectra for TW1, Bridge and LW2 in Figure~\ref{fig:bridge_decompose}. In Figure~\ref{fig:bridge_decompose}, we decompose each blue line in Figure~\ref{fig:spectra} into two components. The green lines represent the radiation coming from the outer emission region and the magenta lines represent those from the inner region.

From Figure~\ref{fig:bridge_decompose}, we can notice that the radiation emitted at a larger $r$, i.e.~from the outer region, will produce a synchrotron peak at a lower energy. This can be explained by the typical synchrotron photon energy given in Equation~(\ref{eq:Esyn}) with $E_{syn}\propto B(r)\sin\varphi(r)\propto r^{-2.5}$. Let us take the Bridge as an example. We choose the representative $r$ to be the midpoint of the emission region from Figure~\ref{fig:phi_r}, i.e.~$1.25R_L$ for the outer region and $0.35R_L$ for the inner region. In this case, the ratio of $r$ is about $3.57$ and hence the characteristic $E_{syn}$ of the inner region is about $24$ times that of the outer region. From the modelled fit, the ratio of the peaks is about $27$. Therefore, the theory matches quite well with the observation in this case.

\section{Conclusion and discussion}
We have used the modified three dimensional outer magnetosphere gap model with the inner boundary of the outer gap being extended from the null charge surface to near the stellar surface. The exact location of the inner boundary does not affect the fitting results. The ``inwardly-extended" part of the outer gap contributes to LW1
and TW2 of the light curve with a slight modification of P1 and P2. Such modified outer gap geometry also plays a vital role in explaining the optical polarization properties of the Crab pulsar \citep{Ta06}. Together with the results of \citet{Ta06} on the gap accelerator, this model explains the pulse profile, phase-resolved spectra and polarization of the Crab pulsar. Also, the outer gap model can explain the observed complex morphology change of light curve as a function of the photon energy \citep{ChTa07}.

Four adjustable parameters are used to simulate the light curve: the inclination angle of the magnetic axis to the rotational axis $\alpha$, $a_1$, the viewing angle also to the rotational axis $\zeta$ and the emission cone pitch angle due to geometry of the emission location $\sin\varphi(R_L)$. As constrained by the phase separation of the double peaks, we choose the values $\alpha=50^\circ$ and $\zeta=76^\circ$. From radio observations, \citet{Ra93} estimated $\alpha\approx 84^\circ$ with $\zeta$ unknown. By using the polarimetric observations at frequencies between $1.4$ and $8.4$ GHz, \citet{MH99} calculated $\alpha\approx 56^\circ$ and $\zeta=117^\circ$. Therefore, further observations is required in order to determine these two parameters more accurately.

For the photon spectra, our model fitting requires $f(R_L)\approx 0.2$, which is larger than the theoretical estimation ($f(R_L)\approx 0.12$) by a factor of $\sim 2$. However, the theory \citep{ZC97,a1} has assumed a vacuum dipole potential. It has been pointed out that in order to explain the observed radiation power on high energy region, current must flow in the accelerator and hence the potential at the gap will be reduced \citep{hirotani06,hirotani07,TaShHiCh06}. Consequently, the gap size with current flow should be larger than the vacuum one.

Moreover, we have taken the stellar radius of a neutron star to be $10^6$ cm. 
Due to the lack of a direct method to determine the size of a neutron star 
\citep{LaPr04}, and the equation of state inside a neutron star given by
 different theoretical models do not provide a unanimous value for 
the neutron star size, we can only determine the magnetic moment, i.e.~$B_pR^3$
, of the pulsar from the energy loss rate. However, although the dipole 
radiation comes from the spin down power of the star, it is not all of it.
For example, the vacuum formula of the dipole radiation for the 
aligned rotator  is usually used to infer the dipole moment of the pulsar. 
On the other hand, MHD solution \citep{Sp06} 
suggested that the effect of the current increases the spindown luminosity 
from the  the vacuum formula. 
 Therefore, we may have overestimated the value of the magnetic moment  
and therefore the strength of the magnetic field. 
Since $E_{syn}\propto B(r)$ hence $B_pR^3$, the synchrotron peak may have been overestimated. Similarly, the peak synchrotron flux $F_{syn}\propto B(r)$ may also be overestimated. Since the power of synchrotron radiation and inverse Compton scattering can be compared by the ratio of the local magnetic energy density and the photon energy density, $\frac{P_{syn}}{P_{ICS}}\sim\frac{U_B}{U_\gamma}\propto\frac{B^2(r)/2\pi}{E_{syn}(r)n_{sy
n}(E_{syn},r)}$ which is independent of $B_pR^3$ shows that $F_{ICS}$ may not be affected. On the other hand, since $F_{syn}$ may have been overestimated, $n_{syn}$ should be smaller. Hence less curvature photons should be absorbed.

\begin{figure}
\includegraphics[angle=90,scale=1.8]{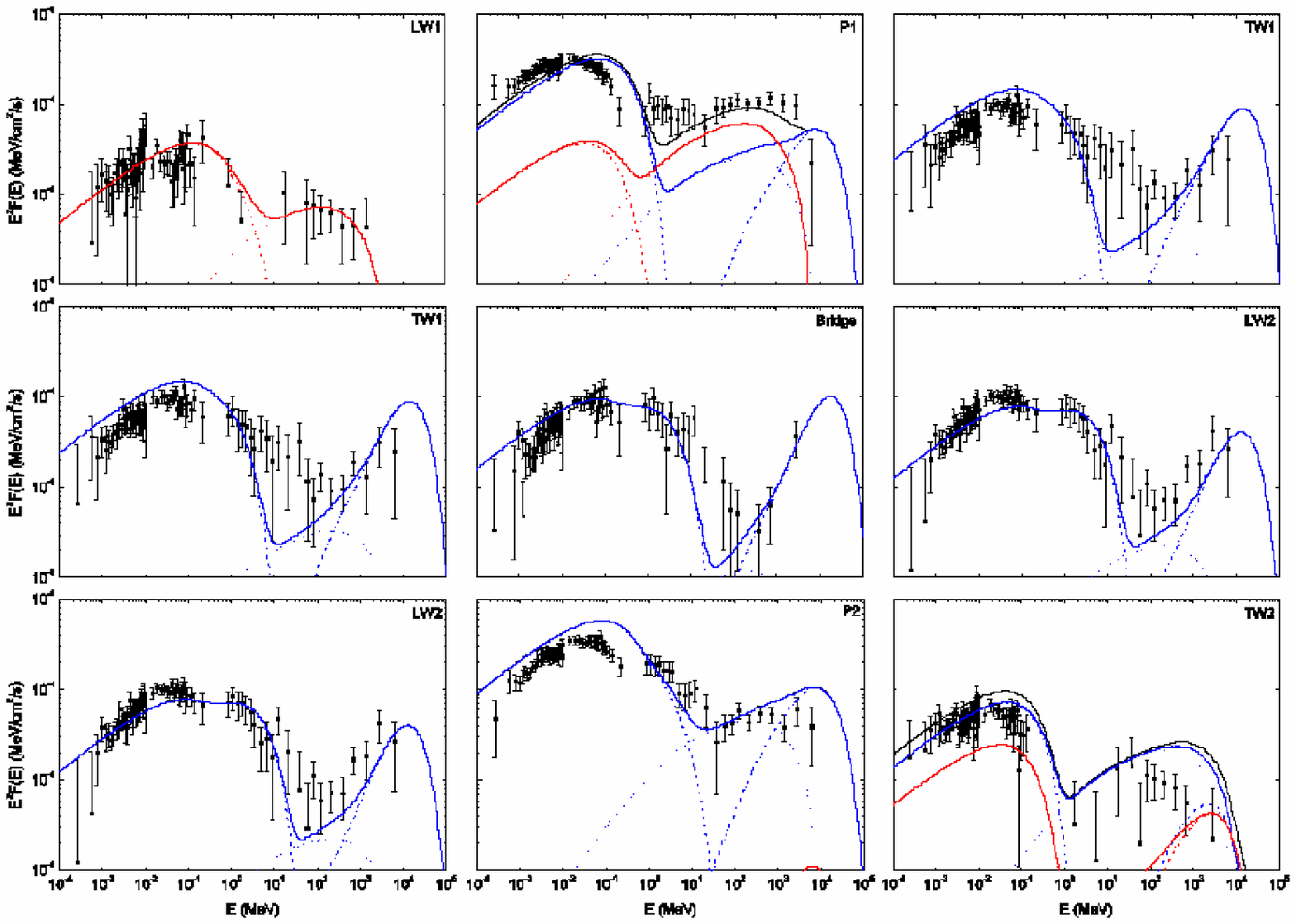}
\caption{Phase resolved spectra of the Crab pulsar from 100 eV to 30 GeV in the 7 narrow pulse-phase intervals. Two spectra (TW1 and LW2) are displayed twice for easy comparison. The observed data are taken from \citet{Ku01}, and the curved lines are calculated by the theoretical model. $\alpha=50^\circ$, $a_1=0.97$ and $\varsigma=76^\circ$. $f(R_L)=0.25$ for the north pole and $f(R_L)=0.22$ for the south pole. The fitting parameters $\sin\varphi(R_L)$ are given in Table~\ref{table:fitting}. \label{fig:spectra}}
\end{figure}

\begin{figure}
\epsscale{.5}
\plotone{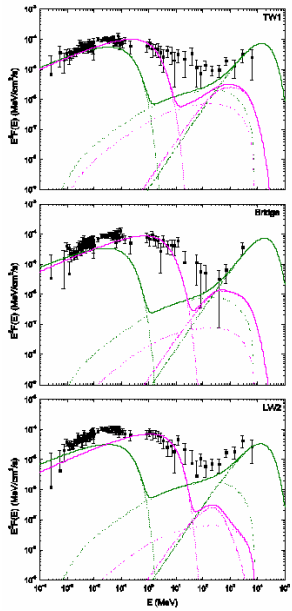}
\caption{The breakdown of the south pole spectra (blue) of the phases TW1, Bridge and LW2 in Figure~\ref{fig:spectra}, each being decomposed into two components according to the emission locations.\label{fig:bridge_decompose}}
\end{figure}

The Crab pulsar is one of the most important pulsars because of its close distance and strong radiation signals in all frequencies. Although observation data of it have been gathered for about 40 years, the underlying physics involved in this pulsar are not completely understood. For example, the structure of the charged particle accelerator in the magnetosphere can be different in the polar cap model, the outer gap model and the newly proposed caustic model \citep{DR03} which proposed a two pole model with the thin gap regions starting from each polar cap to the light cylinder. We hope that further observation by the more sensitive satellites in the future missions, e.g.~GLAST, can bring more crucial and critical information so that we can determine the mechanisms that are truly taking place. For example, as shown in Figure~\ref{fig:spectra}, we predict that there is a clear component ($30$ MeV - $30$ GeV), which is mainly dominated by the survival curvature photons, in TW1, Bridge and LW2. Our model also suggests that the high energy turn over in TW1, Bridge and LW2 is $\sim 10$ GeV, whereas the high energy turn over of LW1 is only $\sim 100$ MeV (much lower than the other phases). This difference results from the consequence of the radiation in LW1 being emitted from a different pole in the region between the inner boundary and the null charge surface, where the local electric field is much weaker (cf.~Eq.~\ref{eq:Er}). With an increase in sensitivity and a widening of the energy range than EGRET, GLAST may be able to confirm our predictions.



\acknowledgments

We are grateful to K. Hirotani, M. Ruderman, S. Shibata and L. Zhang for useful 
discussion and comments, and the anonymous
referee for helpful comments to improve the paper. We also thank L. Kuiper, W. Hermsen 
and D. Thompson for providing the observed data. This work is supported by a RGC grant of 
Hong Kong SAR Government and the National Science Council Excellence Project program in 
Taiwan administered through grant number NSC 96-2752-M007-001-PAE (J.T.).

\clearpage



\clearpage








\clearpage

\clearpage

\clearpage

\clearpage

\clearpage
\begin{table}
\begin{center}
\caption{Phase component definitions for the Crab pulsar.\label{table:Kuiper}}
\begin{tabular}{llccl}
\tableline
Component & Abbreviation & Phase interval & \multicolumn{2}{c}{Width}\\\tableline
Leading Wing 1 & LW1 & $8.8^\circ-30.4^\circ$ & $21.6^\circ$ & ($0.06$)\\
Peak 1 & P1 & $30.4^\circ-66.4^\circ$ & $36^\circ$ & ($0.10$)\\
Trailing Wing 1 & TW1 & $66.4^\circ-102.4^\circ$ & $36^\circ$ & ($0.10$)\\
Bridge & Bridge & $102.4^\circ-142^\circ$ & $39.6^\circ$ & ($0.11$)\\
Leading Wing 2 & LW2 & $142^\circ-167.2^\circ$ & $25.2^\circ$ & ($0.07$)\\
Peak 2 & P2 & $167.2^\circ-206.8^\circ$ & $39.6^\circ$ & ($0.11$)\\
Trailing Wing 2 & TW2 & $206.8^\circ-239.2^\circ$ & $32.4^\circ$ & ($0.09$)\\
Off Pulse & OP & $239.2^\circ-8.8^\circ$ & $129.6^\circ$ & ($0.36$)\\\tableline
\end{tabular}
\end{center}
\end{table}

\clearpage
\begin{table}
\begin{center}
\caption{Summary of fitting parameters $\sin\varphi(R_L)$. We have taken $f(R_L)$ to be $0.25$ for the north pole and $0.22$ for the south pole.\label{table:fitting}}
\begin{tabular}{lcc}
\tableline
Phases & \multicolumn{2}{c}{$\sin\varphi(R_L)$}\\
 & North Pole & South Pole \\\tableline
LW1 & $0.02$ & $-$ \\
P1 & $0.03$ & $0.08$ \\
TW1 & $-$ & $0.08$ \\
Bridge & $-$ & $0.08$ \\
LW2 & $-$ & $0.06$ \\
P2 & $0.08$ & $0.08$ \\
TW2 &  $0.07$ &  $0.08$ \\\tableline
\end{tabular}
\end{center}
\end{table}

\end{document}